\pdfoutput=1
\documentclass[final,1p,times]{elsarticle} 

\usepackage{graphicx}
\usepackage{amssymb} 
\usepackage{amsthm} 

\journal{Nuclear Physics A} 

\begin{document}

\begin{frontmatter} 

\title{The STAR Dilepton Physics Program}

\author{Frank Geurts (for the STAR\fnref{col1} Collaboration)}
\fntext[col1] {A list of members of the STAR Collaboration and
  acknowledgments can be found at the end of this issue.}
\address{Rice University, Houston, TX 77005, USA}

\begin{abstract} 
Dileptons provide ideal penetrating probes of the evolution of
strongly-interacting matter. With the Time-of-Flight upgrade, STAR
is in a unique position to provide large-acceptance dielectron
measurements. We discuss preliminary dielectron results for Au$+$Au
collisions at $\sqrt{s_\mathrm{NN}}=$19.6 - 200~GeV, and compare to
recent model calculations. With upcoming detector upgrades STAR will
further improve its dielectron measurements as well as include dimuon
measurements.
\end{abstract} 

\end{frontmatter}

\section{Introduction}
Experimental results from heavy-ion collisions have long established
the creation of a new state of hot, dense, and strongly interacting
matter \cite{whitepapers}. Throughout the evolution of such a system,
lepton pairs arise from virtual photons, $\gamma^*\rightarrow
l^+l^-$, with its sources varying as a function of the kinematics.
Leptons have a very low cross section with the strongly interacting
medium. Thus, with negligible final state interactions dileptons
provide an excellent penetrating probe of the evolution of a heavy-ion
collision.  A typical dilepton invariant mass spectrum involves a
plethora of sources, ranging from Dalitz decays in the low
invariant mass range to Drell-Yan pair production out at higher invariant
masses. In between, and in addition to these, we find lepton pairs
from vector meson and open heavy-flavor decays, as well as the
possible opportunity to detect thermal radiation emitted from a quark-gluon
plasma.

The dilepton invariant mass spectra allow us to make divisions that
give us an almost chronological view on the evolution of a system. The
prominent sources of dileptons in the low mass range (LMR; $M_{ll} <$
1.1 GeV/$c^2$), in addition to the Dalitz decays, are the direct
decays of the $\rho(770)$, $\omega(782)$, and $\phi(1020)$ vector
mesons, generated after the chemical freezeout. The spectral line
shape of these mesons may reveal imprints of in-medium modifications
possibly related to chiral symmetry restoration. Here, the $\rho$
meson is of special interest given that in thermal equilibrium its
yield dominates that of the $\omega$ and $\phi$ mesons
\cite{RappWambachHees}, and given its relatively short lifetime which
is less than the expected lifetime of the system. In the intermediate
mass range (IMR; 1.1 $< M_{ll} <$ 3 GeV/$c^2$), the production of
dileptons is closely related to the thermal radiation of the
Quark-Gluon Plasma (QGP).  At higher center-of-mass energies, however,
this signal competes with significant contributions from open
heavy-flavor decays such as $\mathrm{c}\bar{\mathrm{c}}\rightarrow
e^+e^-X$, where such charm contributions may get modified by the
medium, too. Heavy-flavor decays continue to contribute into the high
invariant mass region (HMR; $M_{ll}>3$ GeV/$c^2$) in addition to the
primordial dileptons from initial hard scattering between the partons
of the colliding nuclei, as described by the Drell-Yan processes.
Dileptons from the decay of the heavy quarkonia such as the $J/\psi$
and $\Upsilon$ mesons provide means to study deconfinement effects in
the hot and dense medium.

Elliptic flow measurements are used to characterize the azimuthal
asymmetry of momentum distributions. Dilepton elliptic flow
measurements as a function of $p_\mathrm{T}$ have been proposed as an
independent measure to study the medium properties \cite{Chatterjee}.
The combination of certain transverse momentum and invariant mass
ranges allows for different observational windows on specific stages
of the expansion.  Thus, dileptons can be used to further probe the
early stages after a collision and possibly constrain the QGP
equation of state.

In 2010, the STAR experiment at RHIC completed its installation of the
Time-of-Flight (TOF) detector \cite{startof}. The TOF detector brings
large-acceptance and high-efficiency particle identification that not
only extends the hadron identification reach to higher momenta, but
also significantly improves the electron identification in the low
momentum range. This, combined with the RHIC Beam Energy Scan program
in 2010 and 2011, puts STAR in a unique position to measure dielectron
spectra in the low and intermediate mass ranges from top RHIC down to
SPS center-of-mass energies. In these proceedings we present new STAR
results from dielectron measurements at the RHIC top energy of
200~GeV, as well as first results from measurements at several
lower RHIC beam energies. The STAR LMR measurements are compared with
recent model calculations. We conclude these proceedings with an
outlook on the future STAR dilepton program.

\section{Electron Identification and Background Reconstruction}
The electron identification for the results reported in the next
sections involves the STAR Time Projection Chamber (TPC) and the TOF
detector. The TPC detector is the central tracking device of the STAR
experiment and provides reconstructed particle tracks and momenta. The
energy-loss measurements, dE$/$d$x$, in the TPC are used for particle
identification. The TOF detector, with full azimuthal coverage at
mid-rapidity, extends the particle identification range to higher
momenta. The combination of the TOF velocity information and the TPC
energy loss allows for the removal of slower hadrons which contaminate
the electron sample. With a beam-energy specific selection window and
a single track momentum threshold of $p_\mathrm{T}>0.2$~GeV/$c$, the
electron purity in the minimum bias Au$+$Au analysis is 95\%.

\begin{figure}[ht]
\begin{minipage}[t]{0.45\linewidth}
\centering
\includegraphics[width=\textwidth,keepaspectratio]{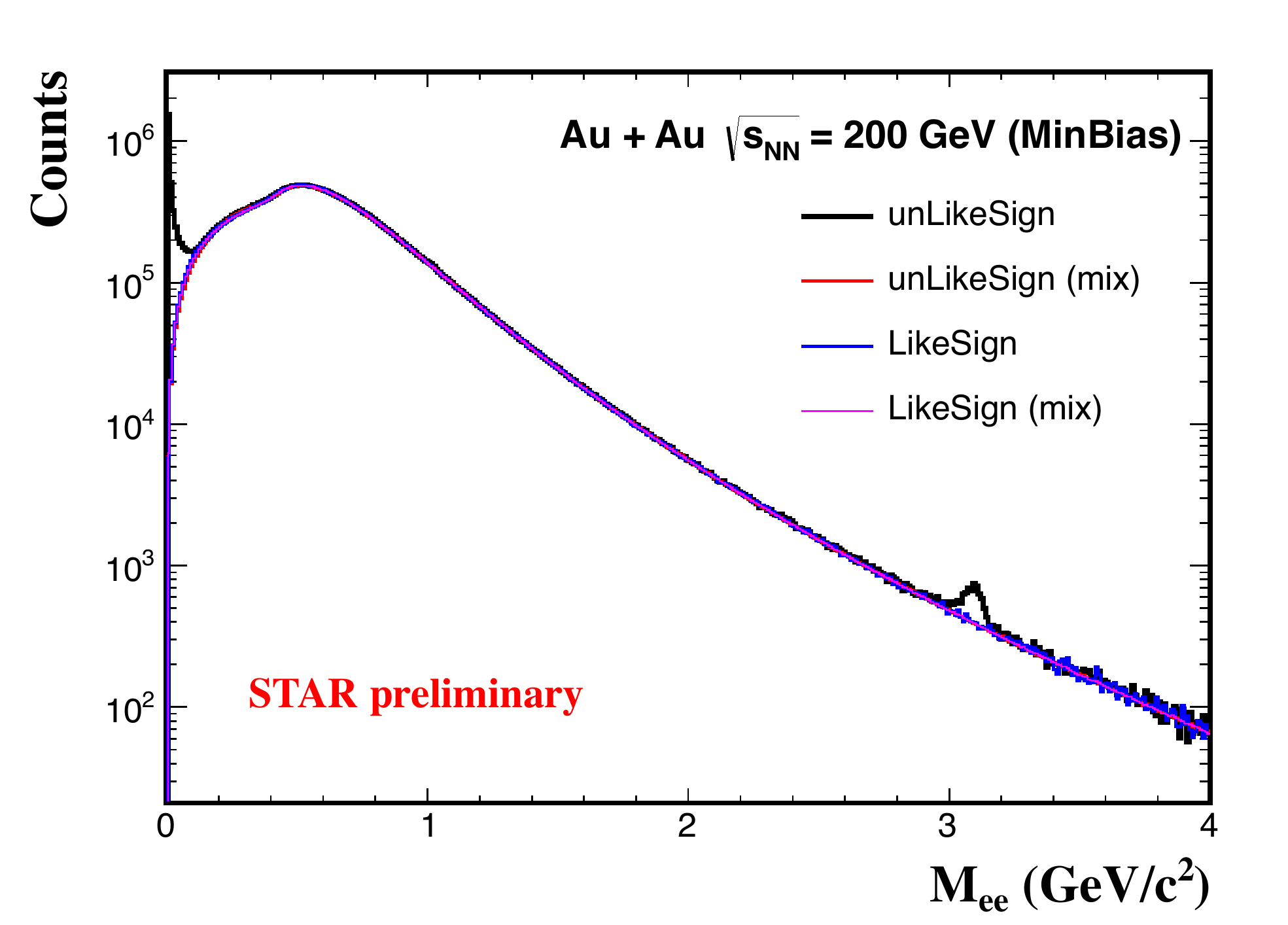}
\caption{(Color online) Unlike-sign, like-sign, and mixed-event
  distributions in Au$+$Au minimum bias collisions at
  $\sqrt{s_\mathrm{NN}}=200$~GeV.}
\label{fig:backgroundAuAu}
\end{minipage}
\hspace{0.05\linewidth}
\begin{minipage}[t]{0.5\linewidth}
\centering
\includegraphics[width=\textwidth,keepaspectratio]{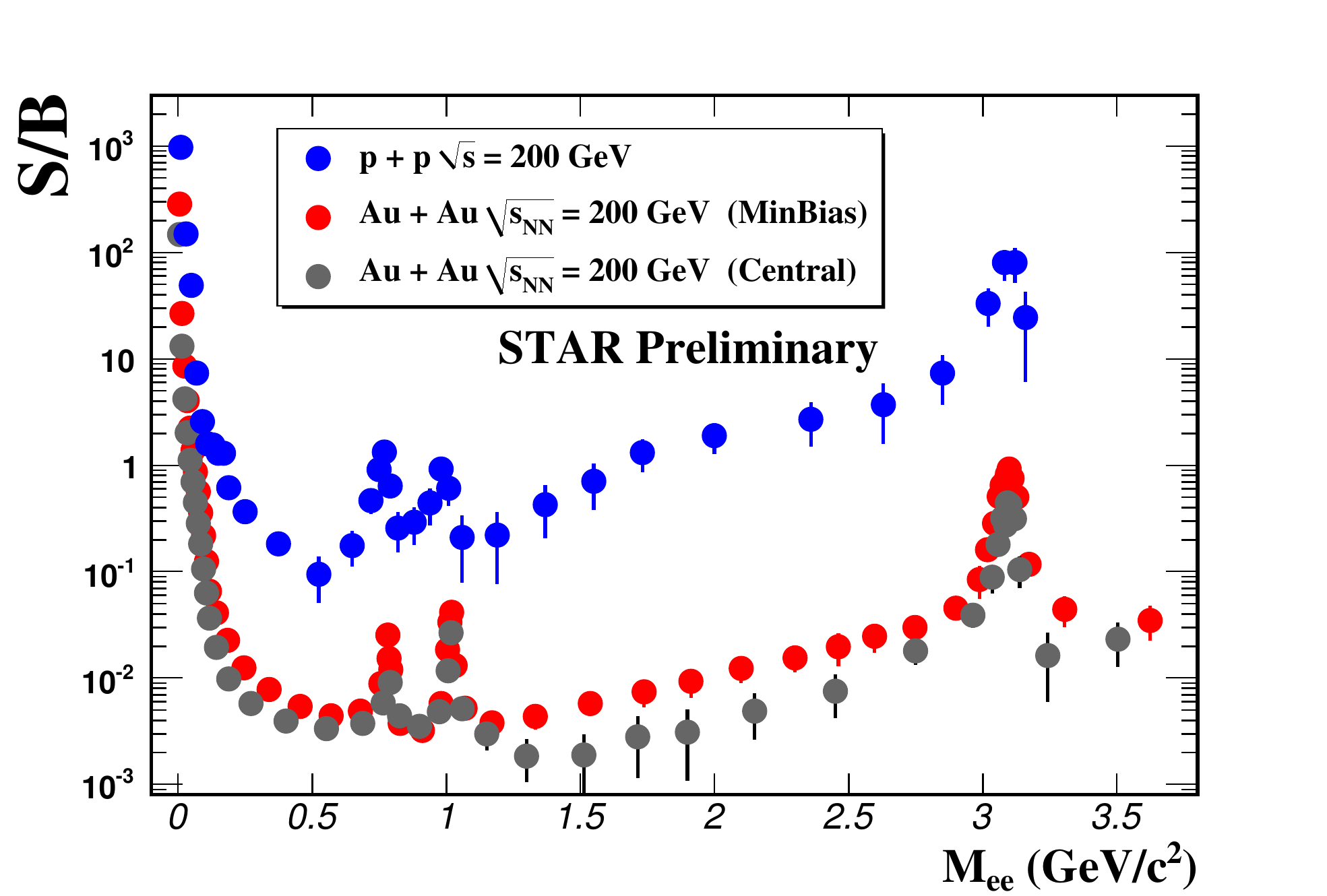}
\caption{(Color online) Signal-to-background ratios in p$+$p, and
  Au$+$Au central, minimum bias events at
  $\sqrt{s_\mathrm{NN}}=200$~GeV \protect\cite{starAuAuQM11}.}
\label{fig:sbratio}
\end{minipage}
\end{figure}

The unlike-sign invariant mass distributions, which are reconstructed
by combining electrons and positrons from the same event, contain both
signal and background contributions. Especially in high-multiplicity
events, the contribution of the combinatorial background is
substantial, see Fig.\ \ref{fig:backgroundAuAu}. The analyses that are
presented in these proceedings use two different methods to determine
the background. 

The mixed-event method combines electrons and positrons from different
events. The events are categorized according to the total particle
multiplicity, the event vertex location along the beam axis, and the
event plane angle. While the statistical accuracy of the background
description can arbitrarily be improved by involving more events, the
mixed-event method fails to reconstruct correlated background sources.
At the lower invariant masses, such correlated pairs arise from jets,
double Dalitz decays, Dalitz decays followed by a conversion of the
decay photon, or two-photon decays followed by the conversion of both
photons \cite{starCocktail}. A background determination in which like-sign
pairs of the same event are combined is able to account for such
correlated contributions. Its drawback, however, is that the
statistical accuracy is only comparable to the unlike-sign, {\em i.e.}
the original raw mass spectrum. Moreover, the like-sign method will
need to consider detector acceptance differences, in contrast to the
(unlike-sign) mixed-event method.

The like-sign method is applied in the in Au$+$Au LMR for
$M_\mathrm{ee}<$750~MeV/$c^2$ at $\sqrt{s_\mathrm{NN}}=$ 200~GeV, for
$M_\mathrm{ee}<$900~MeV/$c^2$ for 39~GeV and 62.4~GeV, and throughout
the full mass range in 19~GeV \cite{HuangQM12,starAuAuQM11}. The
signal-to-background ratio is shown in Fig.\ \ref{fig:sbratio}.

\section{Dielectron Measurements at $\sqrt{s_\mathrm{NN}}$=200~GeV}
STAR recently published the results on its dielectron measurements in
p$+$p at $\sqrt{s}=$200~GeV \cite{starpp}. These results were based on
107 million p$+$p events taken in 2009, with only a partially
commissioned TOF system. The agreement between the measured
yields and the expected yields from a range of hadronic decays,
heavy-flavor decays, and Drell-Yan production, provided an important
test of the analysis methods.

\begin{figure}[ht]
\centering
\includegraphics[width=0.8\textwidth,keepaspectratio]{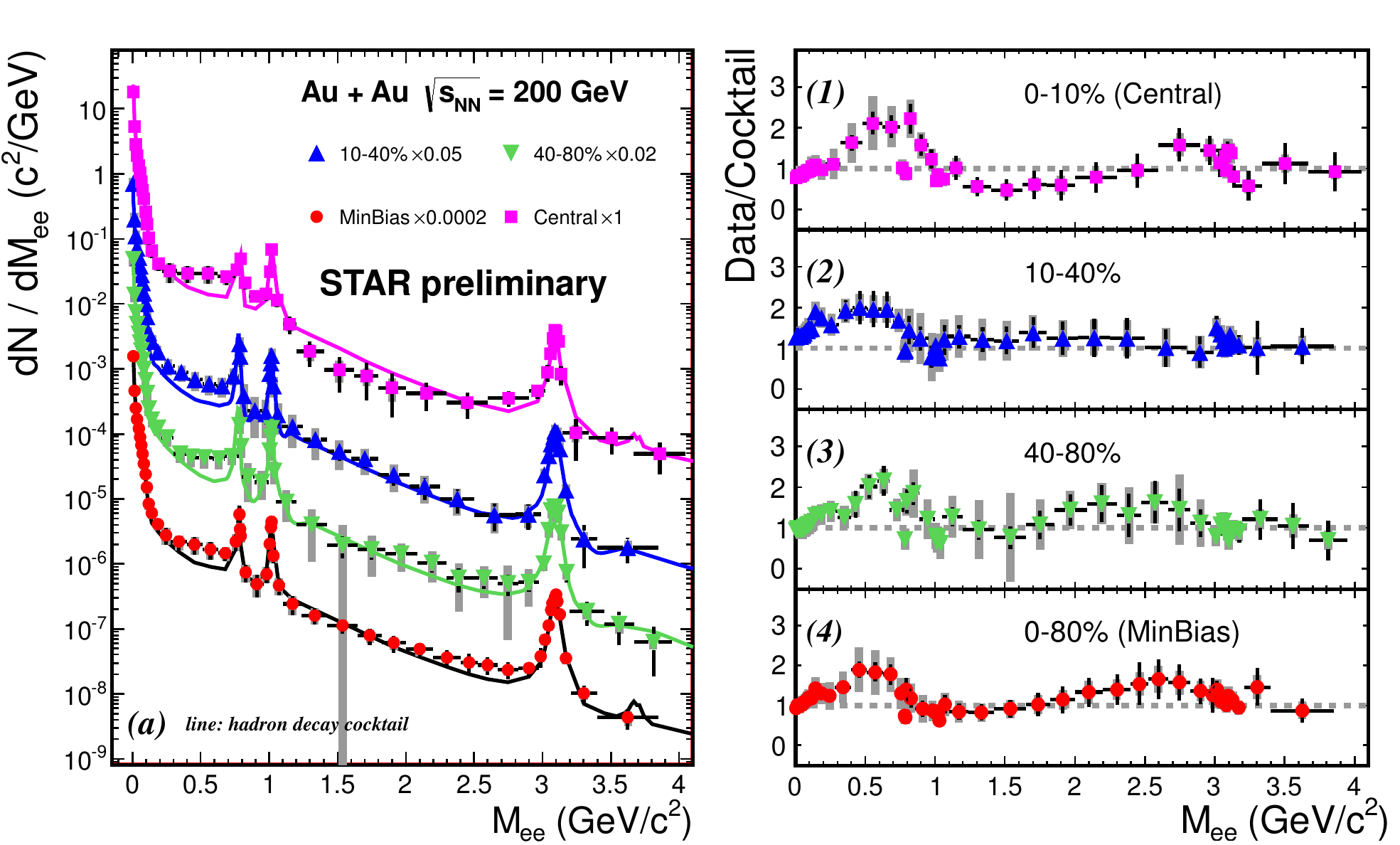}
\caption{(Color online) Dielectron invariant mass spectra for Au$+$Au collisions at
  $\sqrt{s_\mathrm{NN}}=$200~GeV for different centrality selections
  (left panel) and the ratio of data to cocktail (right
  panel). The systematical uncertainties are indicated by boxes.}
\label{fig:dielectronSpectra200GeV}
\end{figure}

Previously at this conference, STAR presented preliminary results of
the LMR and IMR invariant mass spectra in both central and minimum
bias Au$+$Au collisions at $\sqrt{s_\mathrm{NN}}$=200~GeV
\cite{starAuAuQM11}. In the left panel of Fig.\
\ref{fig:dielectronSpectra200GeV}, we now present for the same system
the centrality dependence of the invariant mass spectra in the STAR
acceptance ($|y_{ee}|<1.0$, $|\eta_e|<1$, and $p_\mathrm{T}>
0.2$~GeV/$c$). The measured yields are compared to a cocktail
simulation of expected yields where the hadronic decays include the
leptonic decay channels of the $\omega$, $\phi$, and J$/\psi$ vector
mesons, as well as the Dalitz decays of the $\pi^0$, $\eta$, $\eta'$
mesons. The input distributions to the simulations are based on
Tsallis Blast-Wave function fits to the invariant yields of the
measured mesons \cite{starCocktail}. These functions are used as the
input distributions for the {\sc Geant} detector simulation using the
full STAR geometry.  The $\rho$ meson contributions have not been
included in the cocktails as it may be sensitive to in-medium
modifications which are expected to affect its spectral line shapes
\cite{starrho}. In the IMR, the c$\bar{\mathrm{c}}$ cross section is
based on {\sc Pythia} simulations scaled by the number of
nucleon-nucleon collisions \cite{starCharm}.

\begin{figure}[ht]
\begin{minipage}[t]{0.47\linewidth}
\centering
\includegraphics[width=\textwidth,keepaspectratio]{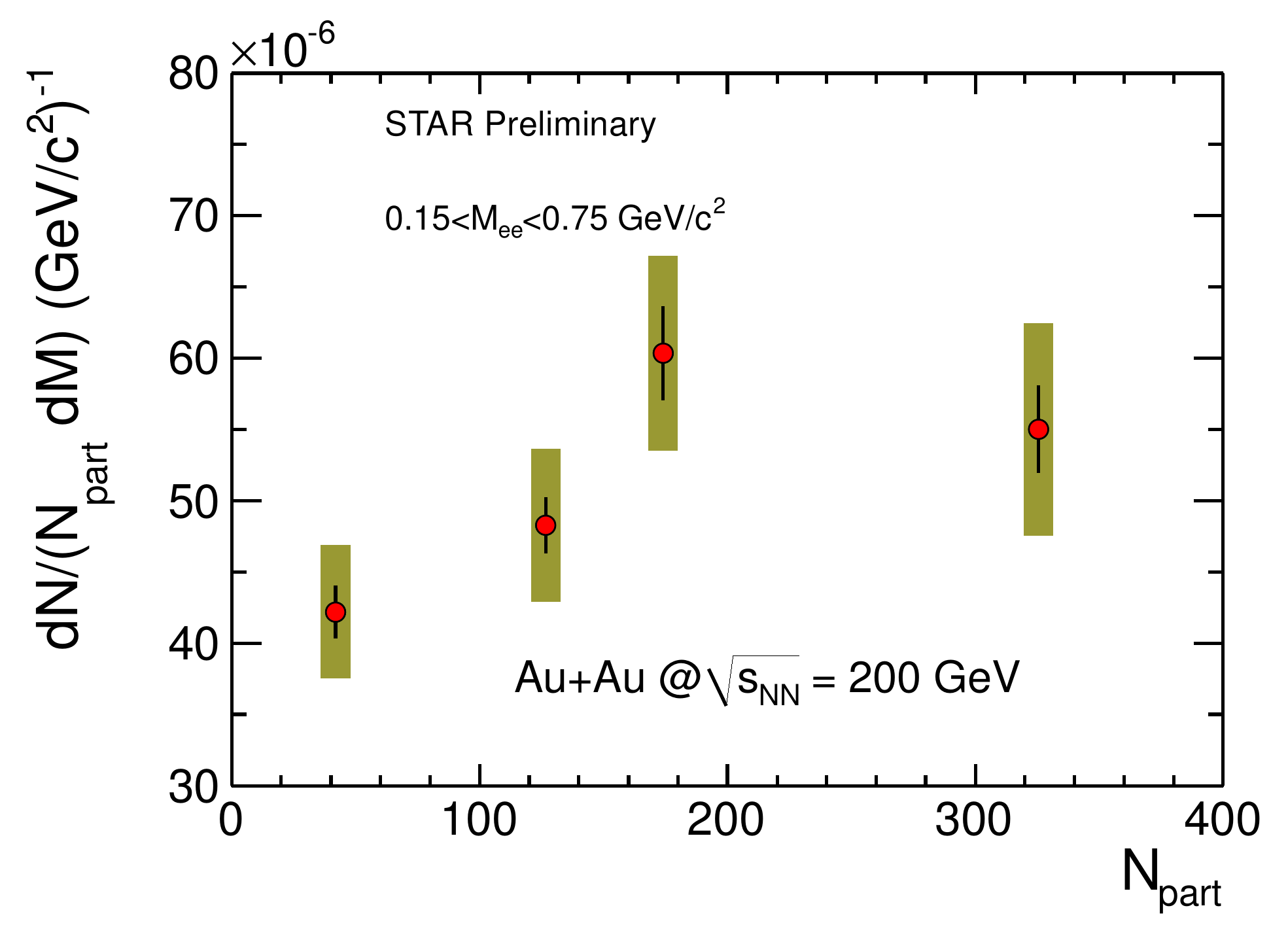}
\caption{Dielectron LMR enhancement scaled by $N_\mathrm{part}$ versus
  centrality ($N_\mathrm{part}$) for Au$+$Au collisions at
  $\sqrt{s_\mathrm{NN}}=$200~GeV \cite{RuanQM12}. The boxes indicate
  the systematical uncertainty.}
\label{fig:LMRenhancement200GeV}
\end{minipage}
\hspace{0.06\linewidth}
\begin{minipage}[t]{0.47\linewidth}
\centering
\includegraphics[width=\textwidth,keepaspectratio]{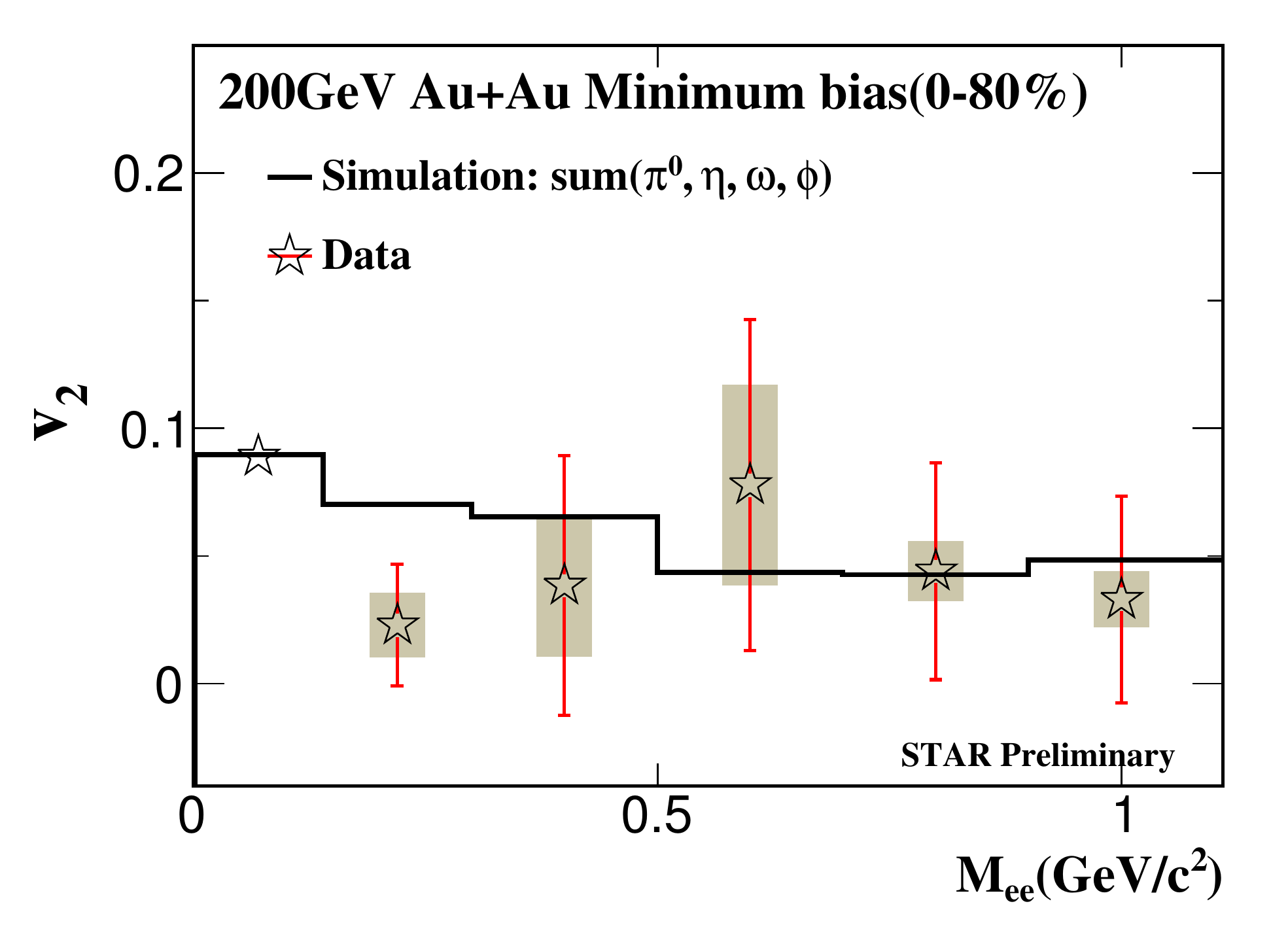}
\caption{Dielectron elliptic flow as a function of the invariant mass
  in minimum-bias Au$+$Au collisions at
  $\sqrt{s_\mathrm{NN}}=$200~GeV. The boxes indicate the systematic
  uncertainty. The black line is the sum of a simulation which
  involved $\pi^0$, $\eta$, $\omega$, and $\phi$-mesons.}
\label{fig:dielectronflow}
\end{minipage}
\end{figure}

In the right panels of Fig.\ \ref{fig:dielectronSpectra200GeV},
the ratios of the data to cocktail yields have been depicted for different
centrality selections. With respect to the cocktail reference a clear
LMR enhancement can be observed for each centrality selection. With
the cocktail as a reference, only little centrality dependence can be
observed. On the other hand, as can be seen in Fig.\
\ref{fig:LMRenhancement200GeV}, a comparison of the dilepton yield
$dN/dM_{ee}$ in the range of $150 < M_{ee} < 750$~MeV/$c^2$ scaled to
the number of participants, $N_\mathrm{part}$, clearly indicates an
increase of the LMR enhancement with increasing centrality. This
agrees with a similar observation in \cite{CERES1}.

In the IMR we observe the cocktail simulations to overestimate the
data in central collisions. This can indicate a modification of the
charm contribution. However, the observed discrepancy is still
consistent within the experimental uncertainties. Future detector
upgrades will allow STAR to further disentangle the potentially
modified charm contributions and provide improved measurements of the
thermal QGP dilepton radiation.

In Fig.\ \ref{fig:dielectronflow}, we present new STAR dielectron
elliptic-flow measurements in Au$+$Au collisions at
$\sqrt{s_\mathrm{NN}}=$ 200~GeV as a function of the dielectron invariant
mass. The results are based on 700 million minimum bias events from an
analysis that combines data sets from the 2010 and 2011 RHIC runs. The
elliptic flow, $v_2$, is calculated using the event-plane method in
which the event plane has been reconstructed from TPC tracks
\cite{starflow}. The ``signal'' elliptic flow, $v_2^\mathrm{signal}$,
is calculated as follows \cite{HuangSQM11}:
\[
 v_2^\mathrm{total}(M_{ee}) =  v_2^\mathrm{signal}(M_{ee})
 \frac{r(M_\mathrm{ee})}{1+r(M_\mathrm{ee})}  + v_2^\mathrm{bkgd}(M_{ee}) \frac{1}{1+r(M_\mathrm{ee})}, 
\]
where $v_2^\mathrm{total}$ is the flow measurement of all dielectron
candidates, $v_2^\mathrm{bkgd}$ the flow measurement of the dielectron
background, and $r(M_\mathrm{ee})$ the mass-dependent
signal-to-background ratio, depicted in Fig.\ \ref{fig:sbratio}. The
expected $v_2$ from a cocktail simulation based on the contributions
from $\pi^0$, $\eta$, $\omega$, and $\phi$ mesons is within
uncertainties consistent with the measurements.

\begin{figure}[ht]
\centering
\includegraphics[width=0.6\textwidth,keepaspectratio]{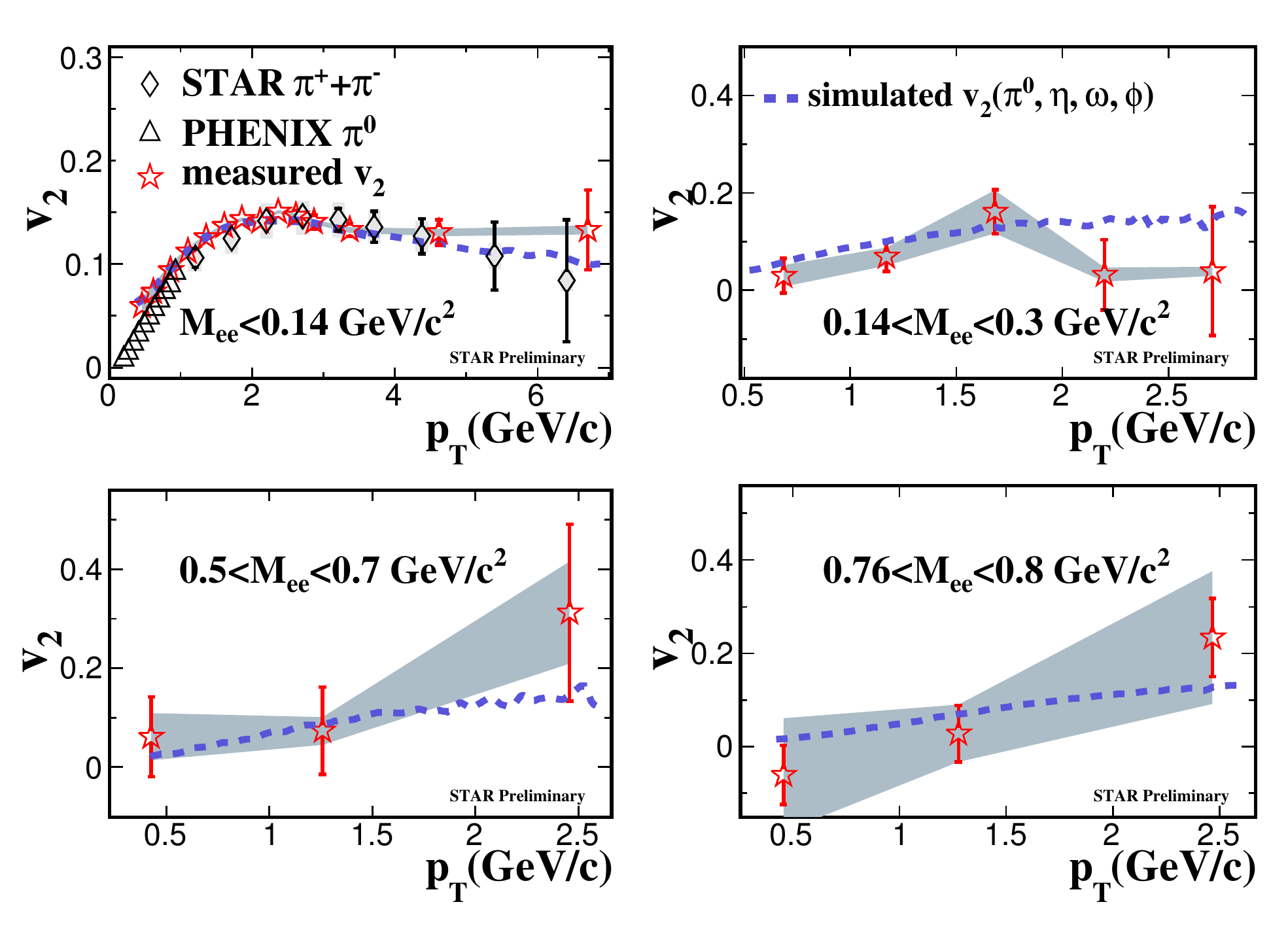}
\includegraphics[width=0.3\textwidth,keepaspectratio]{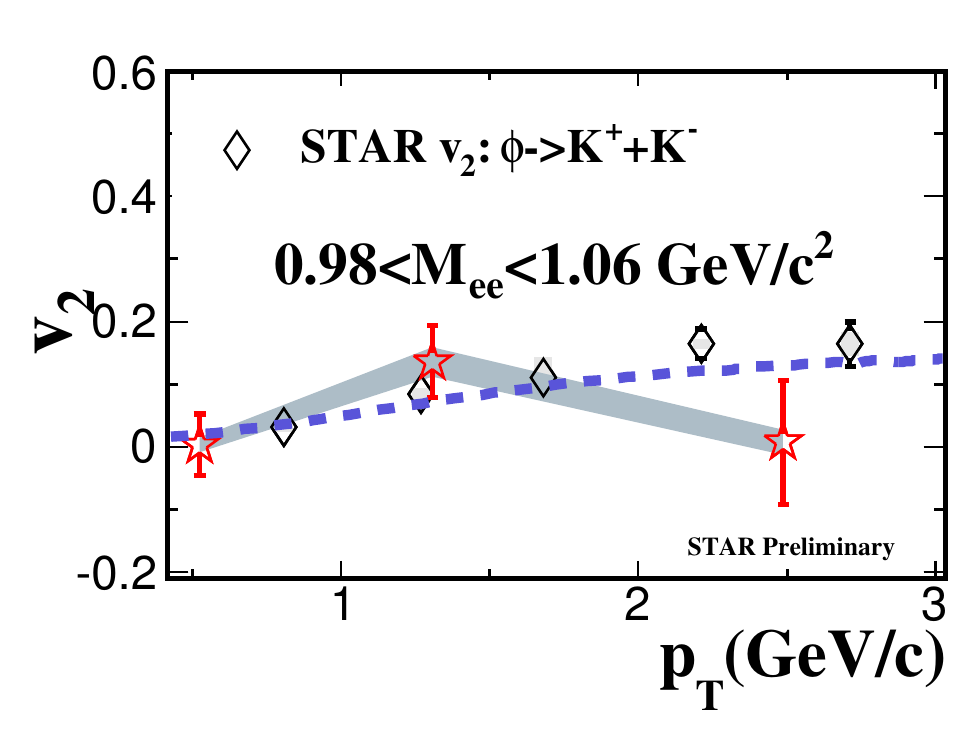}
\caption{Elliptic flow, $v_2$, as a function of $p_\mathrm{T}$ for
  various dielectron invariant mass ranges in Au$+$Au at
  $\sqrt{s_\mathrm{NN}}=$200~GeV (red stars). In addition, in the
  upper left panel the $v_2$ of $\pi^0$ \cite{phenixpi0} and $\pi^\pm$
  mesons \cite{starpiv2} are shown. The $\phi$ meson $v_2$
  measurements in the lower right panel are from STAR. The blue dashed
  lines are the expected $v_2$ from cocktail simulations (see text).
  The grey bands indicate the systematic uncertainties.}
\label{fig:flowpt}
\end{figure}

Differential measurements have been done as a function of centrality
(not shown here) and $p_\mathrm{T}$, see Fig.\ \ref{fig:flowpt}. Both
$v_2(p_\mathrm{T})$ for different dielectron mass windows, and its
centrality dependence in the lower mass bin show a consistency between
the measurements and the simulations. Work is underway to further
extend these measurements into the IMR. It will be important, however,
to disentangle the charm contributions at higher $M_{ee}$. We observe
that the current experimental uncertainties on the STAR data points
are still too large to allow for further constraints on the QGP
equation of state, as is conjectured in e.g. \cite{Chatterjee}.
However, such a sensitivity is experimentally well within reach with
the combination of future data sets, and with an improved understanding
of the charm contributions.

\section{Dielectron Measurements in the Beam Energy Scan}
Measurements performed at top RHIC energies by both STAR and PHENIX
\cite{PHENIX} have shown a clear LMR enhancement which may point to
in-medium modification effects possibly resulting from chiral symmetry
restoration.  Measurements performed at the SPS, by the CERES
\cite{CERES1,CERES2} and NA60 \cite{NA60} collaborations, favor a broadening
of the $\rho$ meson spectral function compared to a dropping-mass
scenario. With the RHIC BES program, STAR is in a unique position to
systematically explore the dilepton production from top RHIC down to
SPS center-of-mass energies.

\begin{figure}[ht]
\begin{minipage}[t]{0.55\linewidth}
\centering
\includegraphics[width=\textwidth,keepaspectratio]{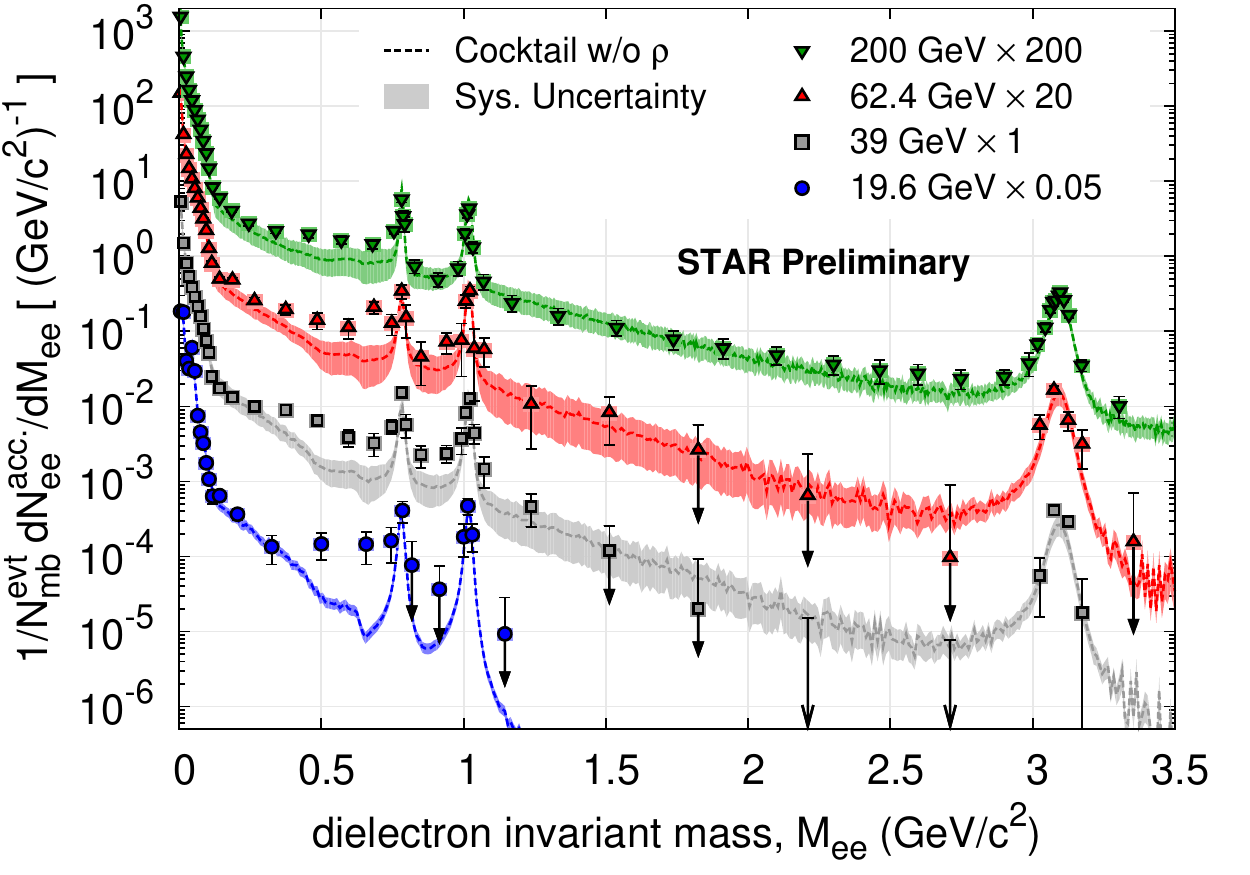}
\caption{(Color online) Background-subtracted dielectron
  invariant-mass distributions from Au$+$Au collisions at
  $\sqrt{s_\mathrm{NN}}=$ 19.6, 39, 62.4, and 200~GeV. The (colored)
  dotted lines show the hadron cocktails (excluding contributions from
  $\rho$ mesons). The (color) shaded areas indicate the systematic
  uncertainties.}
\label{fig:allenergiesdata}
\end{minipage}
\hspace{0.05\linewidth}
\begin{minipage}[t]{0.4\linewidth}
\centering
\includegraphics[width=\textwidth,keepaspectratio]{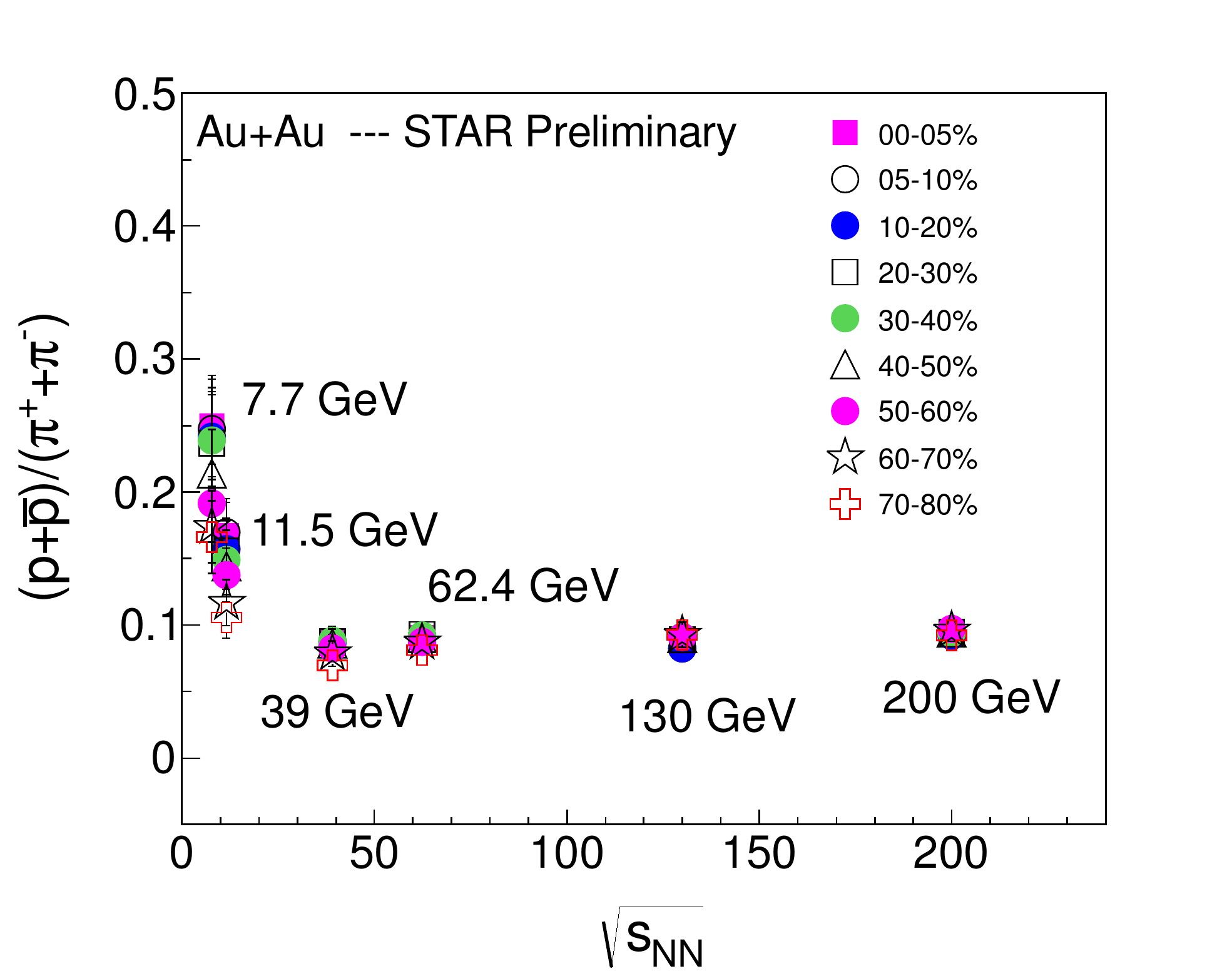}
\caption{$(\mathrm{p}+\bar{\mathrm{p}})/(\pi^+ + \pi^-)$ in Au$+$Au
    collisions as a function of the center-of-mass energy
    $\sqrt{s_\mathrm{NN}}$.}
\label{fig:totalbaryon}
\end{minipage}
\end{figure}

In Fig.\ \ref{fig:allenergiesdata}, the inclusive invariant mass
spectra in Au$+$Au collisions from $\sqrt{s_\mathrm{NN}}=$ 19.6~GeV to
200~GeV are shown, together with the cocktail simulations for each
energy. The cocktail simulations exclude contributions from the $\rho$
meson. For each of the energies we observe a significant LMR
enhancement. While there remains a quantitative discrepancy between
the PHENIX and STAR measurements at $\sqrt{s_\mathrm{NN}}=$ 200~GeV
\cite{PHENIX,geurtsAustin}, we observe a comparable agreement between
the CERES measurements in the Pb$+$Au at
$\sqrt{s_\mathrm{NN}}=$17.2~GeV \cite{HuangQM12,CERES2}.

Various models have shown to be in good agreement with LMR spectrum
measured by STAR in central Au$+$Au collisions at 200~GeV
\cite{PHSD,USTC,Rapp}. The BES measurements provide an excellent
opportunity to further verify robust and consistent model descriptions
down to SPS energies. In-medium broadening of the $\rho$ meson is
expected to be driven by the strong coupling to baryons and thus the
total baryon density since the vector mesons interact symmetrically
with baryons and antibaryons \cite{CERES2,Rapp}.  While the net-baryon
density will vanish at RHIC top energies, the total baryon density
does not change significantly with beam energies down to ~20~GeV, as
can be seen in Fig.\ \ref{fig:totalbaryon}.

\begin{figure}[ht]
\centering
\includegraphics[width=0.8\textwidth,keepaspectratio]{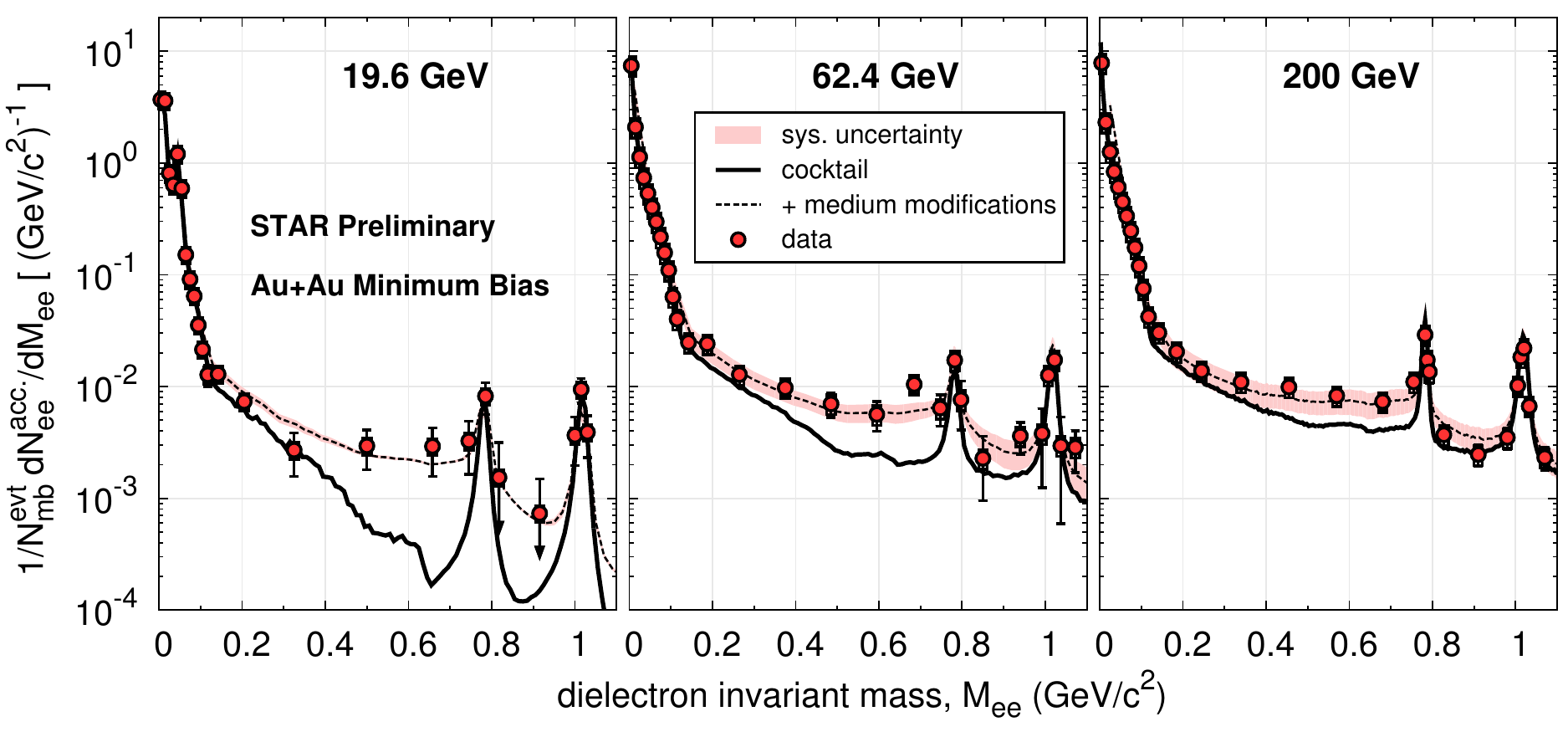}
\caption{BES dielectron invariant-mass distributions in the low
  invariant-mass range from Au$+$Au collisions at 19.6~GeV (left),
  62.4~GeV (middle), and 200~GeV (right panel). The red data points
  include both statistical and systematic uncertainties (boxes). The
  black curve depicts the hadron cocktail, while the dashed line shows
  the sum of the cocktail and model calculations. The latter includes
  contributions from both the HG and the QGP phases.  The systematic
  uncertainty on the former is shown by the light red band.}
\label{fig:inmediumrho}
\end{figure}

In Fig.\ \ref{fig:inmediumrho}, we show the efficiency-corrected
invariant mass spectra for minimum bias Au$+$Au collisions at
$\sqrt{s_\mathrm{NN}}=$ 19.6, 62.4, and 200~GeV, respectively. In each
of the three panels the hadron cocktail simulations include
contributions from Dalitz decays, photon conversions (19.6~GeV only),
and the dielectron decay of the $\omega$ and $\phi$ vector mesons. The
cocktail simulations purposely exclude contributions from $\rho$
mesons. Instead, these are explicitly included in the model
calculations by Rapp \cite{RappWambachHees,Rapp} which involve
in-medium modifications of the $\rho$ meson spectral shape. In this
model a complete evolution of the QGP and thermal dilepton rates in
the QGP and hadron-gas (HG) phases are convoluted with an isentropic
fireball evolution. In the HG phase the $\rho$ ``melts'' when
extrapolated close the conjectured phase transition boundary.
Moreover, it is noted that the top-down extrapolated QGP rates closely
coincide with the bottom-up extrapolated hadronic rates
\cite{RappWambachHees}.

The LMR enhancement measured by STAR can consistently be described by
these model calculations and agrees with a scenario in which the
in-medium modification of the $\rho$ involves a broadening of its
spectral function. Differential studies of the LMR dielectron
distributions as a function of $\sqrt{s_\mathrm{NN}}$, centrality, and
$p_\mathrm{T}$ will allow for a more detailed comparison against these
chiral hadronic models. Further comparisons between these models,
lattice QCD calculations, and experimental data could help provide
explicit evidence of chiral symmetry restoration in heavy-ion collisions.

\section{Summary and Outlook}
Measuring dilepton spectra is a challenging task, where very small
signals typically sit on top of large backgrounds. With the recent TOF
upgrade, STAR has been able to address some of the important physics
questions regarding the LMR enhancement. It has allowed for cross
checks with previous measurements on both ends of the BES energy
spectrum, as well as systematic, large-acceptance measurements at
various other center-of-mass energies. The measurements confirm
observations made at SPS energies in which the LMR enhancement is
attributed to the broadening of the $\rho$ meson. The model
calculations give a robust and consistent picture, and future
differential measurements of the BES LMR data will provide an
excellent tool to further advance the study of chiral symmetry
restoration.

The first years of dielectron measurements at STAR have provided value
input to the field, while a number of open questions still remain. As
the typical production rates for dileptons are rather small, large
event samples are needed to make significantly accurate measurements
at increasingly higher invariant masses. Especially at the lower RHIC
beam energies, the statistical uncertainties are very large (see Fig.\
\ref{fig:allenergiesdata}). The 19.6~GeV measurements have been based
on 28 million Au$+$Au collisions. To achieve statistical uncertainties
at a level of ~10\% and improve our understanding of the baryonic
component to the in-medium effects on the vector mesons, a factor of
10 more statistics would be required.  In order to achieve such an
increase of an order of a magnitude in total luminosity, RHIC will
need the development and installation of electron cooling components.
At top RHIC energy, first results of the LMR dielectron elliptic flow
have been presented. These results are found to be consistent with
expectations from simulations and other elliptic flow measurements.
With an approximate doubling of the currently used combined data
sample and an improved understanding of the IMR charm contributions,
STAR should be able to distinguish between the HG and QGP $v_2$
contributions, and help constrain the latter.

Cocktail simulations indicate that the correlated charm contributions
dominate the IMR dielectron spectra. The {\sc Pythia} simulations that
went into the cocktail simulations assume no medium effects and have
been rescaled to match the charm cross sections to, {\em e.g.},
0.96~mb in 200~GeV Au$+$Au collisions. Both assumptions result in
large uncertainties, while some hints point to a suppresion of the
charm contribution in central collisions when compared to the
minimum-bias data (see Fig.\ \ref{fig:dielectronSpectra200GeV}). These
uncertainties directly affect the extraction of any IMR dilepton
signals related to the thermal QGP radiation. A next round of detector
upgrades \cite{HHuangQM12} will further position STAR to significantly
improve its measurements of the charm contribution at intermediate
dilepton masses. The Muon Telescope Detector (MTD) will enable the
measurement of electron-muon correlations, where both leptons are
measured at mid-rapidity. The dominant source of IMR $e$$-$$\mu$
correlations is c$\bar{\mathrm{c}}$ decay, whereas thermal dilepton
decay will result in $l^+l^-$ pairs only. Consequently, measuring
$e$$-$$\mu$ correlations will be an essential tool for isolating the
IMR thermal contribution.

The primary focus of the MTD physics program \cite{MTD} will be on the
quarkonia measurements at RHIC energies. However, this
large-acceptance muon detector will also allow the STAR dilepton
physics program to encompass dimuon spectra and provide for
measurements of QGP thermal radiation, light vector mesons, and
Drell-Yan production. The MTD is expected to be fully commisioned in
2014.

\end{document}